# An Overview of Portable Distributed Techniques

Sanjay Bansal [1], Nirved Pandey[2],

[1] Dept of CS, RGPV, SRGPGPI
Indore, Madhya Pradesh, 452001, India

[2] Dept of CSE, RGPV, IIPS
Gwalior, Madhya Pradesh, 474011, India

**Abstract**
In this paper, we reviewed of several portable parallel programming paradigms for use in a distributed programming environment. The Techniques reviewed here are portable. These are mainly distributing computing using MPI pure java based, MPI native java based (JNI) and PVM. We will discuss architecture and utilities of each technique based on our literature review. We explored these portable distributed techniques in four important characteristics scalability, fault tolerance, load balancing and performance. We have identified the various factors and issues for improving these four important characteristics.

## 1. Introduction

Computing power of idle hosts is utilized by distributed computing. There are certain strong reasons that justify using distributed computing in comparison than mainframes. Distributed systems offer a better price and performance than mainframes. A distributed system has more total computing power than a mainframe. If one machine crashes, the system as a whole can still survive in distributed system. Thus distributed system provides better reliability. Computing power can be added in small increments in distributed systems. In this way Incremental growth can be achieved. Distributed systems allow many users to access a common database thus provides data sharing. It also allows many users to share expensive peripherals. It makes human-to-human communication easier. It has flexibility, because workload is spread over available machines in most cost effective way. On distributed computing, various heterogeneous devices communicate with each other with different ways.
One of the ways in which Distributing computing can be classified is based on portability; Portable Distributed Computing v/s Non-portable Distributed computing. Within each type, there may be several techniques considered. The basis of the selection a techniques in a group was that these techniques were most prevailing existing distributing computing technique. The other basis of choice for selecting a technique in each group was that these techniques individually represent their group having a distinguished feature. Various distributed computing techniques are:

(1) Distributed Computing through Message Passing without using Java: In this Plain MPI, PVM may be analyzed

(2) Distributed Computing Through Middleware such as COM-DCOM CORBA.

(3) Distributed Computing through .NET Remoting.

(4) JNI: Wrapper: C and C++ codes are interfaced with Java codes using MPI

(5) Pure Java using socket programming and RMI and using these techniques in MPI.

Among these distributing computing techniques, some are portable and some are not portable. In this paper, we have discussed portable techniques mainly distributed computing using pure java based MPI, native java based MPI and PVM.

## 2. Portable Distributed Techniques

A technique is portable if it applies to different versions of the same operating system (e.g., past and future versions of Operating System) and to different operating system

### 2.1 Distributed Computing using Pure Java Based MPI

The Message Passing Interface (MPI) provides a powerful programming paradigm for high performance computing. It is based on the interchange of messages between processes. MPI can be implemented in any programming language because MPI is language independent. Earlier MPI was implemented in C and FORTAN. Performance of C and FORTAN based MPI is high but portability is major drawback. Researchers are devoting efforts to overcome the problem of portability. Portability problem is solved by implementing the MPI with a portable high level language like JAVA. MPI implementation in java has overcome the problem of portability but on the cost of performance [**1**]. For the last few years, both researchers and practitioners have been concerned with the improving the performance of java based MPI.

MPI can be implemented using java in two ways; pure java and impure java implementation. Researchers are trying to solve many issues related to pure java implementation of MPI. These issues are efficient and lightweight distributed computing implementation, design and development of user centric interface for administrator to configure environments, performance evaluation techniques, performance improvements, system interoperatability across a diverse range of devices, efficient implementation, flexibility, fault-tolerance, scalability, transparency, performance with portability, efficient access control, easy to access, high degree of abstraction to ease the programming , load balancing, heterogeneity, malleability(the capability to handle changes in the resources used during a computation) etc[2].

Java is object oriented language with an attractive feature of portability. Hence practitioners are trying to implement a high performance implementation of MPI with different approaches based on MPI standard. In order to implement MPI using java, MPJ is developed. MPJ is a MPI specification for java. It works with impure as well as with pure java [**3**]. Pure java implementations of MPI are two types, Socket based and RMI based. Socket based are flexible but its programming is complex. Programming with RMI is very easy and efficient. Programmers with RMI can make easily distributed application as compared with socket. Another problem with socket is errors. Distributed programming using socket is very difficult as compare to RMI. Hence programming with socket is tough. Programmer can not make distributed system easily with socket [4].

Performance is measured through latency, bandwidth, and total processing time mainly. Performance can further be improved. In our literature we find several java based MPI with specific shortcoming, but no one can full fill all basic requirements. Researchers are also trying to implement "one size fit all". Reason for "one size fit all" is that for some main concern could be portability while for other high bandwidth or low latency. Portability and high performance are contradictory requirements. There is much scope to manage these contradictory requirements. There have been various efforts made by researchers to improve the performance of pure java based MPI, and maintaining the portability. Some researchers implemented MPJ with layered architecture. In this layered architecture java, java new I/O (NIO) based device, a shared memory device are implemented. It brings ability to swap devices at run time and thus manage the contradictory requirements [5].In this layered architecture, one of layer is of native MPI so it adds overheads and the performance improvement is not so significant. There is a strong need to identify the factors affecting performance and explore the ways to improve the performance of pure java based MPI.

JMPI is designed with remote method invocation with object serialization. JMPI is written purely in java. JMPI can be run on any host where java virtual machine (jvm) is available. As compare to mpiJava it is more stable and reliable. Performance is very low as compare to Sun's RMI or KaRMI [6]. It has a good interface tool which can further be improved. Time to time researcher evaluate the performance of all main techniques of distributed paradigm. In their evaluation they found that the Java RMI has higher round trip latency, when compared with MPI and PVM. The reason for higher latency is, in JavaRMI communication is done in terms of objects. When communication is done over the network using Java RMI, the objects are converted into bytes and then these bytes are transmitted over the network. The conversion of objects into bytes is a time consuming job. For this reason, the Java RMI has the highest latency among all paradigms. Java RMI gives the least performance with respect to bandwidth [**7**]. There is a lot of scope to identify the factors, to improve the performance. There is a strong need to make a framework of these critical factors. This framework can provide a better understanding to developers, designers and programmer managers to design and develop the high performance portable distributed system using pure java. Researcher designed JMPI using both RMI and socket [4]. There are further research areas to implement multiple process of an MPI application on same machine as a separate thread which can improve the performance. Best performance is obtained by increasing the number of computers. However latency is not measured for JMPI which can be a research problem for our research work.

## 2.2 Distributed Computing using Impure Java Implementation of MPI

Java native interface implementation of MPI is basically wrapping existing native MPI libraries with the Java. Java Native Interface (JNI) specifies the transition between java code running within the java virtual machine (JVM) and native, system-dependent C or FORTRAN code. One of the first attempts was Bryan Carpenter's mpiJava. mpiJava is an essentially a collection of JNI wrappers to a local C MPI library[8]. It is a hybrid implementation with limited portability. It has to be recompiled against the specific MPI library being used. Researcher developed HPJava to overcome the problem of compiling again [9].

The major research problem is efficient cooperation between java and native codes [10]. JNI also introduce a time overhead and affect the performance. Researcher devoted to measure the time overhead. The overall native method invocation overhead turned out to be about 3-5 times bigger than for ordinary methods [11]. Researcher also worked for a better understanding of JNI-related performance issues that might help developers to make more informed decisions during the software design phase. Still there is much scope to work in same.

JNI performs poorly because most of the data transferred between the virtual machine and the native space must be copied [12]. Researcher avoided the JNI overhead by tightly integrating mpiJava with a version of Hyperion[13].Some researcher suggested direct buffering mechanism to improve the overhead of copying the data[14][ 15].Still performance of native java interface is an issue that need a systematic investigation. There is a need to design a better understanding of all techniques to overcome the shortcomings like performance, limited portability, efficient and effective coordination among native and java codes.

Beside these issues, the mpiJava library is not thread-safe partly because it depends on an underlying native MPI library. Even if the native MPI library was thread-safe, some additional work would be needed to make the mpiJava wrappers thread-safe. There is a tremendous scope to address various issues related with JNI to improve the performance, fault-tolerance, scalability, transparency, load balancing, resource tracking etc. javaMpi is another wrapper around C MPI Library. JNI has another problem of limited portability [3].

With our literature survey we find some key areas related to java native interface. These key areas are distribution of load over different computers, managing resources and scheduling user jobs, fault tolerance techniques improvements, dynamic load balancing [17].There is a tremendous scope to identify the factor and their improvement in stated above.

2.3 Distributed Computing using PVM

PVM system is a distributed programming environment. PVM support heterogeneous distributed computing. PVM is a portable distributed computing. It supports certain forms of error detection and recovery.

The PVM (Parallel Virtual Machine) provides a unified framework .Large systems can be developed in a straightforward and efficient manner. PVM permit a collection of heterogeneous machines on a network to be viewed as a general purpose concurrent computation resource [18]. With our literature survey we found that there are several areas of PVM in which substantial efforts have been done by various researchers. These areas are better technique for dividing and distributing the big task, performance measurement, performance improvement, analysis, load balancing, dynamic resource allocation and utilization, support for the automatic compilation of different object modules for different architectures, fault detection and tolerance, data access, develop software environments that will automatically find a near-optimal mapping for an high computing(HC) program expressed in a machine-independent high-level language.

With PVM, There is a need to divide and distribute a complex big task among various computers. To address this issue, researcher suggested the heterogeneous network computing environment (HeNCE) as a tool. This tool decomposes their application into subtasks. HeNCE also distribute these subtasks to the machines currently available in the HC system [16]. HeNCE allows the programmer to explicitly specify the parallelism for an application by creating a directed graph. HeNCE also has four types of control constructs: conditional, looping, fan out, and pipelining [19].However there is still need for improvement in dynamic environment for task divide and allocation. There is a strong need for dynamic self-adjustment of task deployment and other aspects of self-management. Still there is need to investigate systematically.

From time to time, researchers have been devoted efforts to measure the performance of PVM and compare it with others techniques of distributed computing. PVM has high latency. It is due to daemon. In case of PVM, the daemon is responsible for all communications. It provides flexibility but it eats up the processing power of the machine. Due to this reason the PVM and Java RMI take more time to pass the message (high latency). PVM performance is 2 to 3 times slower than RPC [7]. There is research scope to work to investigate the critical factors and their improvements with respect to performance of PVM.

Researcher also compares PVM with MPI. The total time of execution of the PVM version is about 20% longer than the MPI implementation of the algorithm. PVM environment is built on the concept of virtual machine, which secures three levels of heterogeneity. First, the network level: different networks can create common virtual machine. Second, the machine level: computers with different architecture, operating systems and data formats can create virtual machine. Third, the application level: the architecture of the virtual machine can be matched to the form of a subtask [25]. The PVM environment was created with heterogeneity over performance, while the principal idea of MPI standard is performance and portability over flexibility. There is much scope to explore the ways to make PVM for data parallel application and to improve the performance of PVM.

Performance improvement can also be achieved by reducing the message transmission time. There is much scope to investigate the reasons and factor to reduce message transmission time. Another reason for performance degradation is buffer access time. There is a strong need to make more effort to reduce PVM based memory allocation time, and improve access protocols to reduce overall message buffering time.

PVM does not offer some functionality such as thread safety, multiple communication end-points per task, and default-case direct message routing. In order to solve these issue researchers developed JPVM.JPVM is a java implementation of PVM [26]. JPVM uses direct task-to-task message delivery instead of daemon-to-daemon routed communications. JPVM is based on TCP. However UDP/IP implementation is yet to be suggested and implemented. When performance of PVM, JPVM is compared, we found that performance become worst and bandwidth is also lost [27].There is much scope to address these issues such as thread safety, multiple communication end-points per task, and default-case direct message routing differently rather than implementing in java.

Performance is measured by simulation elapsed time, speedup and bandwidth [7]. Performance may be different for light and heavy traffic. There is much scope to identify the main causes and relationship between load amounts with performance. Researchers have evaluated the significance of reducing buffer access time and message transit time in order to improve the runtime performance. There is a common belief that to reduce the elapsed time of a parallel program, the workload distribution among the processors must be balanced [20]. So there is a research scope in load balancing and distribution among the various computers connected in a distributed network. Many researchers observed that other factors such as inter-processor communication overheads may also cause poor performance. So there is a scope to improve the interprocess communication to improve performance.

By load balancing, performance of PVM can be improved significantly. PVM round robin load balancing is not the correct approach in a heterogeneous industrial environment where different machines have dissimilar performance. As a result, the Applied Research and Technology Lab has developed a new load balancing sub system ,which has the ability to manage PVM spawn taking into account both current CPU load and CPU performance altogether. Tests have been conducted on the new load balance subsystem which results in an increase of PVM performances in a heavy heterogeneous (in terms of workload/performances) environment [21].Results stated there is improvement in performance but there is a need for reasonable improvements. Hence there is a lot of scope to correct the load balancing. There is also need to distribute the load among various machines in order to maximize the usage of machine and minimize the task execution. There is a still need of "intelligent distribution of load".

The main reason for low performance of PVM is daemon thread. It provides the flexibility but eats up all resources. Researcher are also trying to use one of the new general purpose transport protocol called SCTP (Stream Control Transmission Protocol) for direct communications among the tasks [22]. SCTP has been recently standardized by the IETF (Internet Engineering Task Force) and, compared it with TCP (Transmission Control Protocol). It has new interesting characteristics that could be more suitable for parallel applications. Hence there is a lot of scope to explore the ways for this [23].

Fault-tolerance becomes an important requirement in distributed systems. Many researchers are devoting their efforts in development of a fault-tolerant PVM. Researcher presents a transparent, non-blocking check pointing protocol to support fault-tolerance for PVM applications. Even if the applications have dynamic number of processes, the protocol can be implemented as a user-level library and, therefore, the changes in PVM library and operating system is not necessary[24].However due to check point, an overhead is introduced that could make it unsuitable for many application. So there is a need for a low overhead fault-tolerance technique for PVM.

From our extensive literature survey, there is a lot of research scope in improving the overall utilization of resource. It will consequences improvements in performance of PVM.

## 3. Conclusions

Based on our discussion in section two, we derive the several findings about the portable distributed techniques. Performance of java based MPI is low as compared to plain MPI on cost of portability. Portability and performance are contradictory requirement. Performance can be further improved by improving the factors affecting the performance. Conversion of byte codes to machine codes adds an extra overhead on the performance. In case of JNI, by improving the coordination between the java to native code and vice-versa performance can be improved. Daemon eats up all processing power of computing nodes in PVM. Thus by use of the new general purpose protocol, performance of PVM can improved. More adaptive improved and fuzzy algorithm is required to make load balancing more efficient.

## References

[1] V. Getov, S. Flynn-Hummel and S. Mintchev., "High-Performance parallel programming in Java: Exploiting native libraries," In ACM 1998 Workshop on Java for High Performance Network Computing, Palo Alto, February 1998, Concurrency: Practice and Experience, 1998.

[2] G. F. Jordi, "Performance Improvement of Multithreaded Java Applications Execution on Multiprocessor Systems," Computer Architecture Department (DAC), Technical University of Catalonia (UPC), Barcelona (Spain), July 2005.

[3] G. Vladimir, L. Quanming, a M. Thomas and M. Williams, "Message-Passing Computing with Java: Performance Evaluation and Comparisons," 1066-6192/01$10.000, 2001 *IEEE*, pp. 173-177.

[4] B. SeungJun and A. JinHo, "Implementation and Performance Evaluation of Socket and RMI based Java Message Passing Systems," 0-7695-2867-8/07 $25.00, 2007, IEEE, DOI 10.1109/SERA.2007.60, pp. 153-159.

[5] B. Mark, C. Bryan and S. Aamir, "MPJ: Enabling Parallel Simulations in Java," www/acet.reading.ac.uk/projects/mpj/docs/res/DSGTR19062005 .

[6] M. Steven, K. Israel and C. Mani Krishna "JMPI: Implementing the Message Passing Standard in Java," www.ecs.umass.edu/ece/realtime/publications/morin02.pdf.

[7] Q. Kalim and R. Haroon "A Performance Evaluation of Rpc, Java Rmi, Mpi And Pvm," Malaysian Journal of Computer Science, Vol. 18 No. 2, December 2005, pp. 38-44.

[8] M. Baker, B. Carpenter, G. Fox, S. Ko and S. Lim "mpiJava: An Object-Oriented java Interface to MPI," Lecture Notes In Computer Science; Vol.1586, proceedings of the 11 IPPS/SPDP'99 Workshops held in conjunction with the 13th International Parallel Processing Symposium and 10th Symposium on Parallel and Distributed Processing, pp.748-762 ,Year of Publication: 1999, ISBN:3-540-65831-9.

[9] C. Bryan and F. Geoffrey "Hpjava: A Data Parallel Programming Alternative," 1521-9615/03/$17.00, 2003, IEEE.

[10] K. Dawid and S. Vaidy, "Efficient Cooperation between Java and Native Codes - JNI Performance Benchmark," Emory University, Dept of Math and Computer Science.

[11] Z. Bao-Yin, Y. Ze, Y. Guang-Wen and Wei-Min Zheng, "Dynamic Load-Balancing and High Performance Communication in Jcluster," 1-4244-0910-1/07/$20.00 c_2007 IEEE.

[12] B. Pugh and J. Spacco, "MPJava: High-Performance Message Passing in Java using Java.nio," In Proc. 16th Intl. Workshop on Languages and Compilers for Parallel Computing (LCPC'03), LNCS vol. 2958, pp. 323-339, College Station, TX, USA, 2003.

[13] H. Philip and R. Mathew "Cluster computing With Java," Computing in Science and Engineering**,** Volume 7, Issue 2 (March 2009), pp. 34-39, Year of Publication: 2009, ISSN: 1521-9615, 1521-9615/05/$20.00 © 2005 IEEE.

[14] B. Mark, C. Bryan and S. Aamir, "MPJ Express: Towards Thread Safe Java HPC," Cluster Computing, 2006, IEEE International Conference CLUSTER 2006, pp. 1-10.

[15] M.A. Baker, B. Carpenter, and A. Shafi, "MPJ Express: Towards Thread Safe Java HPC," Proceedings of the IEEE International Conference on Cluster Computing (Cluster 2006), Barcelona, Spain, September, 2006, ISSN: 1552-5244.

[16] A. Beguelin, J. Dongarra, A. Geist, R.Manchek and K. Moore, "HeNCE: A Heterogeneous Network Computing Environment," Tech. Rep. CS-93-205 (August 1993), University of Tennessee.

[17] A. Jameela, M. Nader, J. Hong and S. David, "An Agent-Based Infrastructure for Parallel Java on Heterogeneous Clusters," Proceedings of the IEEE International Conference on Cluster Computing (CLUSTER'02) 0-7695-1745-5/02 $17.00 © 2002 IEEE.

[18] V. S. Sunderam, ''PVM: A Framework for Parallel Distributed Computing,'' Concurrency: Practice and Experience, vol. 2, no. 4, pp. 315-339, December 1990.

[19] M. Muthucumaru, D. B. Tracy, and S. Howard Jay , "Heterogeneous Distributed Computing: Encyclopedia of Electrical and Electronics Engineering," J. G. Webster, editor, John Wiley & Sons, New York, NY, 1999,Vol. 8, pp. 679-690.

[20] P. Kovendhan and D. Janakiram, "The Edge Node File System: A Distributed File System for High Performance Computing," Scalable Computing: Practice and Experience, Vol. 10, Number 1, pp. 115–130, http://www.scpe.org.

[21] D. B. Christian, P. Guido , P. Emiliano De, G. Riccardo and G. Francesco, "PVM Advanced Load Balancing in industrial environment," Proceedings of the 14th Euromicro International Conference on Parallel, Distributed and Network-Based Processing (PDP'06), 1066-6192/06 $20.00 © 2006 IEEE.

[22] R. R. Stewart and Q. Xie, "Stream Control Transmission Protocol (SCTP)," A Reference Guide, Addison-Wesley, 2002.

[23] M. Petrone and R. Zarrelli, "Enabling PVM to build Parallel Multidomain Virtual Machines," Proceedings of the 14th Euromicro International Conference on Parallel, Distributed and Network-Based Processing (PDP'06),
1066-6192/06 $20.00 © 2006 IEEE.

[24] N. Thoai and D. V. Hung, "Checkpoint and Recovery for Parallel Applications with Dynamic Number of Processes," Sixth International Symposium on Parallel and Distributed Computing (ISPDC'07), 0-7695-2936-4/07 $20.00 © 2007.

[25] B. Butrylo , F. Musy , L. Nicolas , R. Scorretti and C. Vollaire, "New Trends in Parallel Electromagnetic Fields Computation," Proceedings of the International Conference on Parallel Computing in Electrical Engineering (PARELEC'02), 0-7695-1730-7/02 $17.00 © 2002 IEEE.

[26] D. Thurman, "JavaPVM," available from http://www. isye.gatech.edu/chmsr/JavaPVM/.

[27] K. Dincer, "jmpi and a Performance Instrumentation Analysis and Visualization Tool for jmpi," First UK Workshop, Java for High Performance Network Computing at EuroPar 1998, September 1998, http://www.cs.cf.ac.uk/hpjworkshop/.

**Author's Profile**

**Sanjay Bansal** has passed B.E. (Elex & Telecom Engg.) and M.E. (Computer Engineering) from Shri Govindram Seksariya Institute of Technology and Science, Indore in 1994 and 2001 respectively.

He has worked as a lecturer in Medi-Caps Institute of Technology. Presently he is working as Reader in Shri Gujrati Professional Institute, Indore. He is pursuing Phd from Rajeev Gandhi Proudyogiki Vishvavidyalaya, Bhopal, India. His research areas are load balancing, fault-tolerance, performance and scalability of distributed system.

**Nirved Pandey** passed B.E. (Electrical Engineering) in 1988 from G.E.C. Jabalpur (R.D.V.V.) and M.Tech (Computer Technology) from I.I.T. Delhi in 1996. He has been awarded Phd in 2008 by ABV-IIITM, Gwalior. He has worked as a lecturer, sr. lecturer and then Reader in MITS, Gwalior. Presently he is working as a Principal at IIPS Gwalior. He has published many International and national research papers. He has attended many international conferences. His research areas are distributed system, query optimization etc.